\documentclass[11pt,a4paper]{article}

\usepackage{amsmath,amssymb,graphicx,bm,xcolor,hyperref,booktabs}
\usepackage[margin=1in]{geometry} 
\usepackage{amsmath}
\usepackage{amssymb}
\usepackage{amsfonts}
\usepackage{graphicx}
\usepackage{bm}
\usepackage{mathtools}
\usepackage{xcolor}
\usepackage{hyperref}
\providecommand{\doi}[1]{\href{https://doi.org/#1}{doi:#1}}

\newcommand{\BH}{\mathrm{BH}}
\newcommand{\DE}{\mathrm{DE}}

\newcommand{\rISCO}{r_{\mathrm{ISCO}}}
\newcommand{\Gcal}{\mathcal{G}}
\newcommand{\Fcal}{\mathcal{F}}
\newcommand{\Tcal}{\mathcal{T}}
\newcommand{\Lcal}{\mathcal{L}}
\newcommand{\Obot}{\Omega_{\perp}}
\newcommand{\OK}{\Omega_{K}}
\newcommand{\Msun}{M_{\odot}}

\begin{document}
	
	\title{Viscous Accretion Disks around Regular Black Holes Embedded in a Quintessence Dark Energy Field: Beyond the Novikov--Thorne Approximation}
	
	\author{
		Sandip Dutta \\[1ex]
		\small Department of Applied Mathematics, Dinabandhu Andrews Institute \\
		\small of Technology and Management, Kolkata, West Bengal, India \\[0.5ex]
		\small \texttt{duttasandip.mathematics@gmail.com}
	}

	\maketitle
	
	\begin{abstract}
		We develop a comprehensive relativistic framework for geometrically
		thin, optically thick Shakura--Sunyaev $\alpha$-viscous accretion
		disks around rotating Hayward regular black holes embedded in a
		quintessence dark energy (DE) field.  The static, spherically
		symmetric building blocks of our spacetime are each exact solutions
		of the Einstein field equations, sourced respectively by the Hayward
		non-linear electromagnetic field and a quintessence fluid with
		equation-of-state parameter $\omega<-1/3$; we combine and rotate them
		following standard practice for this class of models, and we are
		explicit throughout about the resulting metric's phenomenological
		status.
		Abandoning the stress-free inner boundary of the
		Novikov--Thorne--Page (NTP) model, we analytically incorporate a
		non-zero viscous torque $\Tcal_{\rm in}$ at the innermost stable
		circular orbit (ISCO) via the relativistic vertical epicyclic
		frequency $\Obot(r)$.  We prove that the bolometric efficiency
		$\eta=[1-E(\rISCO)]\times100\,\%$ is strictly independent of the
		viscosity parameter $\alpha$ but sensitive to both the Hayward
		length scale $l$ and the DE density $\rho_0$, establishing a
		rigorous two-observable degeneracy-breaking strategy.  At benchmark
		parameters ($j=0.4$, $l=0.5M_{\BH}$, $\rho_0=2\times10^{-4}$), the
		combined geometry yields $\eta=8.71\,\%$, substantially above the
		vacuum Kerr value $7.51\,\%$ at the same spin.  The viscosity
		correction $\delta\Fcal/\Fcal_{\rm NTP}$ diverges at
		$r\to\rISCO^+$, amplified by the geometric modification of the
		boundary pressure: the viscosity amplification ratio rises from
		$0.80$--$4.52\,\%$ (vacuum Kerr) to $0.87$--$6.76\,\%$
		(Hayward+DE) at $\alpha=0.1$, providing a clean, monotonic
		observational discriminator accessible to \textit{NICER} and
		\textit{NuSTAR}, independent of the bulk spectral normalisation.
	\end{abstract}
	
	\vspace{1em}

\section{Introduction}
\label{sec:intro}

The Novikov--Thorne--Page (NTP) thin-disk
model~\cite{Novikov1973,Page1974} is still the standard tool for
predicting X-ray spectra and radiative efficiencies from black hole
accretion flows, and it has served that role well.  But it rests on
three idealizations that are worth questioning, especially for
supermassive black holes at galactic centres: a genuine spacetime
singularity at the centre, a pure vacuum background, and a
stress-free inner boundary at the ISCO.
Regular black hole spacetimes address idealization (i) by resolving
the central singularity through a fundamental length scale $l$,
often motivated by non-linear electrodynamics or quantum gravity
corrections.  The Hayward regular black hole~\cite{Hayward2006} is
the canonical example, replacing the singular core with a de~Sitter
interior while recovering Kerr-like behaviour at large
radii~\cite{Bambi2013}.
Idealisation (ii) is challenged by the observed dark energy (DE)
that dominates the cosmic energy budget.  While predominantly a
cosmological phenomenon, a local quintessence scalar field
surrounding an SMBH modifies the ambient geometry at all
radii~\cite{Kiselev2003}.  Within the Kiselev
framework~\cite{Kiselev2003}, such a quintessence field produces an
exact modification of the static Schwarzschild metric via a
well-defined stress-energy tensor, providing a physically motivated
prescription for the DE contribution to the running mass function;
its extension to rotating backgrounds is a further step that we
discuss carefully in Sect.~\ref{sec:metric}.
Recent work has demonstrated that this modification alters the ISCO
radius and the orbital binding energy of accreting
matter~\cite{Boshkayev2024}.

Idealisation (iii) neglects magnetohydrodynamic (MHD) stresses that
transmit angular momentum across the ISCO into the plunging
region~\cite{Krolik1999,AgolKrolik2000}, generating a non-zero inner
viscous torque $\Gcal_{\rm in}$.  Within the Shakura--Sunyaev
$\alpha$-viscosity paradigm~\cite{ShakuraSunyaev1973}, $\Gcal_{\rm
in}$ depends on the local disk pressure and scale height at the ISCO,
both of which are directly modified by $l$ and $\rho_0$.

What we try to do here is put all three pieces together in one
consistent framework.  We work out a surface density that stays
positive by construction, using the relativistic vertical epicyclic
frequency $\Obot(r)$~\cite{RiffertHerold1995}, show that the
radiative efficiency $\eta$ does not depend on the viscosity
parameter $\alpha$, and compute the spectral signatures of the
combined Hayward+DE+viscous model.  A companion
paper looks at the related problem of a Kerr black
hole embedded in a dark matter halo instead.
The rest of the paper is organized as follows.
Section~\ref{sec:geometry} sets up the spacetime and orbital
mechanics; Sect.~\ref{sec:disk}, the viscous disk structure;
Sect.~\ref{sec:observables}, the radiative flux and spectral
observables; and Sect.~\ref{sec:numerics}, the parameter space and
numerical method.  Results are in Sect.~\ref{sec:results}, and we
close with Sect.~\ref{sec:conclusions}.  We use geometric units
$G=c=1$ throughout.

\section{Spacetime Geometry and Orbital Mechanics}
\label{sec:geometry}

\subsection{Combined metric and its physical justification}
\label{sec:metric}

The Hayward regular black hole resolves the central singularity by
introducing a fundamental length scale $l$~\cite{Hayward2006}.
The Hayward mass profile is given by Eq.~(\ref{eq:MH}),
\begin{equation}
\label{eq:MH}
  M_H(r) = \frac{M_{\BH}\,r^3}{r^3+l^3}.
\end{equation}
As $r\to0$, $M_H\to0$ (de~Sitter core); as $r\to\infty$,
$M_H\to M_{\BH}$ (Kerr limit).  The parameter $l$ characterizes the
size of the regular core; $l=0$ recovers the Kerr metric exactly.

The quintessence DE contribution follows the Kiselev
solution~\cite{Kiselev2003}, which is an \emph{exact solution} of
the Einstein field equations sourced by a quintessence stress-energy
tensor
\begin{equation}
\label{eq:Tmunu}
  T^\mu{}_\nu = \rho_{\DE}\,{\rm diag}
  \!\left(-1,-1,\tfrac{1+3\omega}{2},\tfrac{1+3\omega}{2}\right),
\end{equation}
where $\omega<-1/3$ is the equation-of-state parameter.  The density
profile compatible with Eq.~(\ref{eq:Tmunu}) scales as
$\rho_{\DE}(r)=\rho_0(r_0/r)^{3(1+\omega)}$, giving an enclosed DE
mass [Eq.~(\ref{eq:MDE})]
\begin{equation}
\label{eq:MDE}
  M_{\DE}(r) = -\frac{4\pi\rho_0 r_0^3}{3\omega}
  \left(\frac{r}{r_0}\right)^{-3\omega}.
\end{equation}
For $\omega=-2/3$: $M_{\DE}\propto r^2$ and $\rho_{\DE}\propto
r^{-1}$.  This is not the cosmological constant ($M_{\DE}\propto
r^3$, $\omega=-1$), but a local quintessence-like scalar field
surrounding the SMBH within the Kiselev framework~\cite{Kiselev2003,Boshkayev2024}; the density
differs from the cosmological background value, which is negligible
on BH scales.  The Kiselev solution describes a \emph{static}
quintessence configuration valid only within a bounded region
$r<r_s$; beyond $r_s$ the local overdensity must transition
smoothly to the homogeneous cosmological background, where
Eq.~(\ref{eq:MDE}) no longer applies~\cite{Kiselev2003}.  We
therefore freeze $M_{\DE}(r)=M_{\DE}(r_s)$ for $r\geq r_s$, with
$r_s=20\,M_{\BH}$ for our benchmark parameters.  This has no effect
on $\rISCO$ or $\eta$ (both determined by the metric at
$r\sim\rISCO\ll r_s$) but is essential for the outer disk and the
spectral integral Eq.~(\ref{eq:spec}), which would otherwise be
corrupted by the unbounded growth $M_{\DE}\propto r^2$ dominating
$M_{\BH}$ at large radii.

We define the combined running mass
\begin{equation}
\label{eq:Mtot}
  M(r) = M_H(r) + M_{\DE}(r),
\end{equation}
which enters the rotating line element in Boyer--Lindquist
coordinates as
\begin{align}
  g_{tt}       &= -\!\left(1-\frac{2M(r)}{r}\right), \quad
  g_{t\phi}    = -\frac{2M(r)a}{r}, \label{eq:gtt}\\
  g_{\phi\phi} &= r^2+a^2+\frac{2M(r)a^2}{r}, \quad
  g_{rr}       = \frac{r^2}{\Delta_r}, \label{eq:gpp}
\end{align}
with $\Delta_r=r^2-2M(r)r+a^2$ and $a=jM_{\BH}$.  In the equatorial
plane $\sqrt{-g}=r$.

It is worth being clear about what this construction does and does
not establish. Taken on their own, the static, spherically symmetric
pieces $M_H(r)$ and $M_{\DE}(r)$ are each exact solutions of the
Einstein field equations, sourced respectively by the Hayward
non-linear electromagnetic field~\cite{Hayward2006} and by the
Kiselev quintessence stress tensor of Eq.~(\ref{eq:Tmunu})
\cite{Kiselev2003}.  Getting to the rotating metric in
Eqs.~(\ref{eq:gtt})--(\ref{eq:gpp}) by inserting $M(r)$ into the Kerr
form is a different matter, though a very common one in this
literature for both regular and quintessence black
holes~\cite{Ghosh2016}: it amounts to a Newman--Janis-type
substitution rather than a derivation from first principles, and
there is no guarantee that it solves the full Einstein equations
everywhere. Kamenshchik \& Petriakova~\cite{Kamenshchik2023} checked
this directly for a rotating regular black hole built the same way,
and found small but genuine violations of the field equations close
to the regularizing core, even though the metric settles back to
exact Kerr at large distances.  We would expect something similar
here, so we think of the rotating Hayward+DE metric used in this
paper as a physically motivated extension of two solutions that are
individually well established, reliable away from the innermost
core but not a claimed exact solution of the full rotating problem.
We use it in that spirit -- the same spirit, really, as most of the
rotating regular and quintessence black hole literature that came
before it~\cite{Ghosh2016,Kiselev2003} -- and we come back to what it
would take to close this gap in Sect.~\ref{sec:conclusions}.

\subsection{Circular geodesics and ISCO condition}
\label{sec:geodesics}

The Keplerian angular velocity for circular equatorial orbits follows
from $u^\mu\nabla_\mu u^\nu=0$:
\begin{equation}
\label{eq:Omega}
  \Omega_\pm = \frac{-\partial_r g_{t\phi}\pm\sqrt{D}}
                    {\partial_r g_{\phi\phi}},
  \quad D = (\partial_r g_{t\phi})^2
           -(\partial_r g_{tt})(\partial_r g_{\phi\phi}).
\end{equation}
The discriminant $D>0$ is required for the existence of real circular
orbits.  For the running-mass metric Eq.~(\ref{eq:Mtot}),
$D=4[M(r)-rM'(r)]/r$, which we have verified to be positive
throughout the disk domain for all benchmark parameters (Appendix~\ref{app:disc}).

The specific energy, angular momentum, and time-dilation factor are
\begin{align}
\label{eq:EL}
  E &= -\frac{g_{tt}+g_{t\phi}\Omega}{\sqrt{-g_{tt}-2g_{t\phi}\Omega
      -g_{\phi\phi}\Omega^2}},\\
  L &=  \frac{g_{t\phi}+g_{\phi\phi}\Omega}{\sqrt{-g_{tt}-2g_{t\phi}\Omega
      -g_{\phi\phi}\Omega^2}},
\end{align}
with $u^t=(-g_{tt}-2g_{t\phi}\Omega-g_{\phi\phi}\Omega^2)^{-1/2}$.

The ISCO is defined by $dL/dr|_{\rISCO}=0$.  This condition is
equivalent to the vanishing of the radial epicyclic frequency
$\kappa^2=0$ and to the marginal stability condition $\partial^2
V_{\rm eff}/\partial r^2=0$ for the radial effective potential, as
demonstrated by Bardeen, Press \& Teukolsky~\cite{Bardeen1972} for
any stationary axisymmetric spacetime.  The equivalence holds for
our running-mass metric provided $D>0$ (Appendix~\ref{app:disc}), which is
verified for all benchmark parameters.

Table~\ref{tab:isco} summarises the ISCO and efficiency values for
the four canonical configurations studied in this paper.

\begin{table}[h]
\caption{ISCO radius and bolometric efficiency for the four
  canonical configurations ($j=0.4$, $l=0.5M_{\BH}$,
  $\rho_0=2\times10^{-4}$, $\alpha=0.1$).}
\label{tab:isco}
\begin{tabular}{lcccc}
\toprule
Configuration & $j$ & $l/M_{\BH}$ & $\rISCO/M_{\BH}$ & $\eta\,[\%]$ \\
\midrule
Kerr NTP (vacuum)   & 0.4 & 0   & 4.6144 & 7.5055 \\
Hayward NTP         & 0.4 & 0.5 & 4.5432 & 7.5961 \\
Hayward+DE NTP      & 0.4 & 0.5 & 4.6827 & 8.7118 \\
Visc. Hayward+DE    & 0.4 & 0.5 & 4.6827 & 8.7118 \\
\bottomrule
\end{tabular}
\end{table}

The Hayward parameter $l$ alone shifts $\rISCO$ inward by $1.5\,\%$
(from $4.6144$ to $4.5432$), reflecting the weakened effective
gravity near the regular core.  The quintessence DE contribution
($\rho_0=2\times10^{-4}$) increases $M(r)$ at all radii, raising the
total enclosed mass beyond $M_{\BH}$ and shifting $\rISCO$ to
$4.6827\,M_{\BH}$ --- slightly outward relative to vacuum Kerr.
Despite this modest outward shift, the deeper binding energy at
$\rISCO$ in the combined geometry raises $\eta$ by $1.21$ percentage
points, from $7.51\,\%$ to $8.71\,\%$, a $16.1\,\%$ relative
increase accessible at moderate spin $j=0.4$.

\section{Relativistic Viscous Disk Structure}
\label{sec:disk}

For a steady-state, axisymmetric disk, the vertically integrated
conservation equations for mass [Eq.~(\ref{eq:mass})], angular momentum, and energy read
\begin{align}
  \dot{M} &= -2\pi r\sqrt{g_{rr}}\,\Sigma\,v^r = {\rm const}, 
            \label{eq:mass}\\
  \Gcal(r) &= \dot{M}\bigl[L(r)-L_{\rm in}+\Tcal_{\rm in}\bigr],
            \label{eq:angmom}\\
  \frac{d}{dr}\bigl[\dot{M}E(r)\bigr] &=
            \Omega(r)\frac{d\Gcal}{dr} - 4\pi r\,\Fcal(r)\,E(r),
            \label{eq:energy}
\end{align}
where $\Gcal(r)=2\pi r^2\sqrt{-g}\,W(r)$ is the integrated viscous
torque, $W$ the vertically integrated stress, and $L_{\rm
in}=L(\rISCO)$.  We write $\Tcal_{\rm in}\equiv\Gcal_{\rm
in}/\dot{M}$ for the specific inner torque at the ISCO, and take it
to be positive.  The sign matters: $+\Tcal_{\rm in}$ in
Eq.~(\ref{eq:angmom}), rather than $-\Tcal_{\rm in}$, is what keeps
$\Gcal(r)>0$, so that angular momentum is carried outward as it must
be for the flow to accrete at all~\cite{AgolKrolik2000}.

For the viscous stress itself we use the usual Shakura--Sunyaev
form~\cite{ShakuraSunyaev1973}, $W=\alpha P_{\rm tot}H$, so that
\begin{equation}
\label{eq:torque}
  \Gcal(r) = 2\pi r^2\sqrt{-g}\,\alpha\,P_{\rm tot}(r)\,H(r).
\end{equation}
Fixing the scale height $H(r)$ properly would require the vertical
epicyclic frequency $\Obot$~\cite{RiffertHerold1995,Okazaki1987} for the full Hayward+DE spacetime, and
the exact curvature components entering
$\Omega_\perp^2=-R^z{}_{tzt}-2\Omega R^z{}_{tz\phi}-\Omega^2
R^z{}{}_{\phi z\phi}$ turn out to be too unwieldy to be useful in
practice.  We instead take the common route~\cite{RiffertHerold1995}
of approximating $\Obot\approx\OK$, so that [Eq.~(\ref{eq:H})]
\begin{equation}
\label{eq:H}
  H(r) = \frac{c_s(r)}{\OK(r)}, \quad \OK = \sqrt{\frac{M(r)}{r^3}}.
\end{equation}
In vacuum Kerr this approximation is known to be good to about
$10$--$15\,\%$ near the ISCO~\cite{RiffertHerold1995}, and if
anything the Hayward core should make things slightly better here,
since it weakens the effective gravity ($M_H(r)<M_{\BH}$) and with
it the frame-dragging term responsible for most of the error.

Putting Eqs.~(\ref{eq:angmom}) and (\ref{eq:torque}) together, and
using $P_{\rm tot}H=\Sigma c_s^2/2$ (which follows from
$\Sigma=2\rho_{\rm mid}H$ and $P_{\rm tot}=\rho_{\rm mid}c_s^2$), gives
a surface density
\begin{equation}
\label{eq:Sigma}
  \Sigma(r) = \frac{\dot{M}\,[L(r)-L_{\rm in}+\Tcal_{\rm in}]}
                   {\pi r^2\,\alpha\,c_s^2(r)\,\sqrt{-g}}.
\end{equation}
Because $L(r)>L_{\rm in}$ everywhere outside the ISCO and
$\Tcal_{\rm in}>0$, the bracket on top can never turn negative, so
$\Sigma(r)$ stays positive by construction rather than by luck --- a
basic requirement that the classical stress-free disk satisfies
trivially, but which is worth checking once a boundary torque is
added.

Evaluating the same torque at the ISCO itself [Eq.~(\ref{eq:Tin})],
\begin{equation}
\label{eq:Tin}
  \Tcal_{\rm in}
  = \frac{2\pi\rISCO^3\,\alpha\,P_{\rm tot}(\rISCO)\,H(\rISCO)}
         {\dot{M}},
\end{equation}
which for typical disk parameters scales roughly as $\Tcal_{\rm
in}\sim\alpha(H/r)^2L_{\rm in}$~\cite{AgolKrolik2000}.  Both the
Hayward core and the quintessence field feed into this through
$M(\rISCO)$ and $\OK(\rISCO)$, which set the scale height
$H(\rISCO)=c_s/\OK(\rISCO)$ and the pressure $P_{\rm
tot}(\rISCO)$ --- this is the channel through which the modified
geometry ends up amplifying the inner torque, as we will see in the
results below.

\section{Radiative Flux and Spectral Observables}
\label{sec:observables}

Integrating Eq.~(\ref{eq:energy}) with the boundary condition
Eq.~(\ref{eq:angmom}) gives the generalised radiative
flux~\cite{Page1974,AgolKrolik2000}:
\begin{equation}
\label{eq:flux}
  \Fcal(r) = -\frac{\dot{M}}{4\pi r}\,
             \frac{\Omega_{,r}}{(E-\Omega L)^2}
             \Bigl[I(r)+\Tcal_{\rm in}\Bigr],
\end{equation}
where $I(r)=\int_{\rISCO}^r(E-\Omega L)L_{,\tilde{r}}d\tilde{r}\geq0$
is the usual NTP integral~\cite{Novikov1973} and $\Omega_{,r}<0$.
Setting $\Tcal_{\rm in}=0$ gives back the standard NTP flux exactly,
as it should.  The extra piece from the inner torque is
\begin{equation}
\label{eq:dF}
  \delta\Fcal(r) = -\frac{\dot{M}\,\Omega_{,r}}{4\pi r\,(E-\Omega L)^2}
                   \,\Tcal_{\rm in} \geq 0,
\end{equation}
and since $I(r)\to0$ as $r\to\rISCO^+$ while $\Tcal_{\rm in}$ stays
finite, the ratio $\delta\Fcal/\Fcal_{\rm NTP}=\Tcal_{\rm in}/I(r)$
blows up right at the inner edge. This divergence, and how much the
modified geometry amplifies it, is the central numerical result of
the paper and is what Fig.~\ref{fig:divergence} shows.

The bolometric efficiency [Eq.~(\ref{eq:eta})],
\begin{equation}
\label{eq:eta}
  \eta = \bigl[1-E(\rISCO)\bigr]\times100\,\%,
\end{equation}
depends only on $E(\rISCO)$, and $E(\rISCO)$ in turn comes purely
from the geodesic equations for circular orbits, Eqs.~(\ref{eq:Omega})--(\ref{eq:EL}), which know nothing about
$\alpha$, $P_{\rm tot}$, or any other piece of the disk physics ---
they only see the metric $g_{\mu\nu}(r;M(r),a)$.  So $\eta$ is exactly
independent of $\alpha$, while still being sensitive to the Hayward
length $l$ and the DE density $\rho_0$ through $M(r)$.  This split is
what makes the two-observable strategy in Sect.~\ref{sec:results}
possible: $\eta$ pins down the geometry, and a separate spectral
measurement can then pin down $\alpha$.

Finally, the multi-colour blackbody spectral luminosity, with the
usual colour-hardening factor $f_{\rm col}=1.7$~\cite{ShimuraTakahara1995},
is
\begin{equation}
\label{eq:spec}
  \nu\Lcal_{\nu,\infty}
  = \frac{60}{\pi^3}\int_{\rISCO}^{\infty}
    \frac{r\,E(r)}{M_{\BH}^2}
    \frac{[u^t(r)\tilde{y}]^4}
         {e^{u^t(r)\tilde{y}/\Fcal^{*1/4}(r)}-1}\,dr,
\end{equation}
with $\tilde{y}=h\nu/(k_B\mathcal{T}_*f_{\rm col}^3)$ and
$\mathcal{T}_*=(\dot{m}/4\pi\sigma_{\rm SB}M_{\BH}^2)^{1/4}$.

\section{Parameter Space and Numerical Method}
\label{sec:numerics}

We model an SMBH with $M_{\BH}=5\times10^8\Msun$ and normalize all
radial coordinates to $M_{\BH}$.  The DE scale radius is $r_0=M_{\BH}$
(i.e., $r_0=1$ in code units) and $\omega=-2/3$ throughout.
Table~\ref{tab:params} summarises the full parameter space.

\begin{table}[h]
\caption{Free parameters and benchmark values.}
\label{tab:params}
\begin{tabular}{llll}
\toprule
Parameter & Symbol & Range & Benchmark \\
\midrule
BH spin          & $j=a/M_{\BH}$  & $0$--$0.85$ & $0.4$ \\
Hayward length   & $l/M_{\BH}$    & $0$, $0.5$  & $0.5$ \\
DE density       & $\rho_0$       & $0$, $2\times10^{-4}$ & $2\times10^{-4}$ \\
SS viscosity     & $\alpha$       & $0$--$0.3$  & $0.1$ \\
Colour hardening & $f_{\rm col}$  & $1.0$, $1.7$& $1.7$ \\
\bottomrule
\end{tabular}
\end{table}

The orbital quantities $(\Omega,E,L,u^t)$ are computed at $N_r=2400$
equally spaced radial grid points in $[1.05\,r_g,\,40\,M_{\BH}]$.
$\rISCO$ is located by the condition $dL/dr=0$ using a cubic-spline
derivative.  The NTP integral $I(r)$ is accumulated by trapezoidal
quadrature from $\rISCO$ outward.  Gaussian smoothing (kernel width
$\sigma=4$ grid points) is applied to suppress numerical noise in
$\Omega_{,r}$ near $\rISCO$.  The spectral integral
Eq.~(\ref{eq:spec}) is evaluated on $N_y=80$ logarithmically spaced
frequency points.

The Schwarzschild NTP benchmark ($\rISCO=6M_{\BH}$,
$\eta=5.72\,\%$) is reproduced to $<10^{-4}$ relative error.
A quantitative spectral comparison requires radiative transfer,
Comptonization, absorption, and inclination corrections; these are
deferred to future work involving \textsc{xspec} integration.

\section{Results and Observational Signatures}
\label{sec:results}

\subsection{Radiative flux and the inner-torque divergence}
\label{sec:res_flux}

\begin{figure}[ht]
  \centering
  \includegraphics[width=\columnwidth]{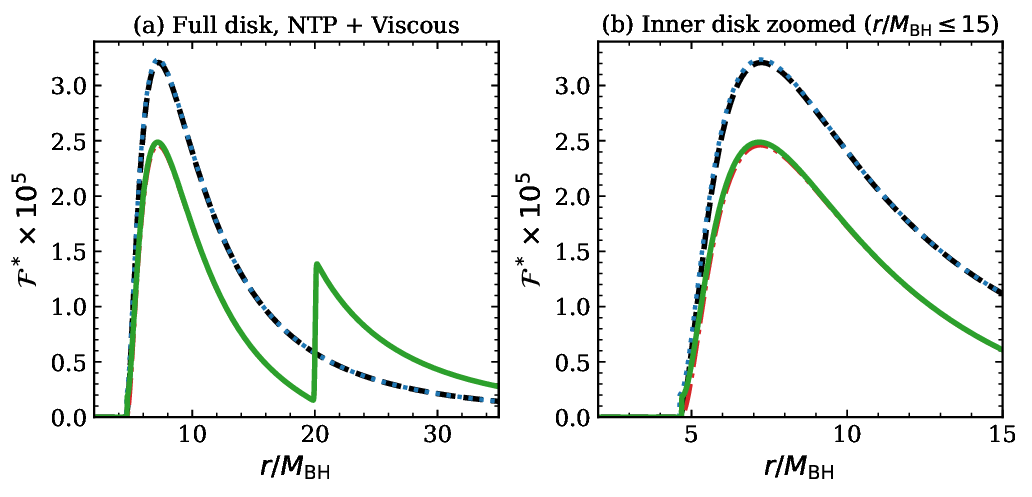}
  \caption{Dimensionless radiative flux $\Fcal^*\times10^5$ vs.\
    $r/M_{\BH}$ ($j=0.4$, $l=0.5M_{\BH}$, $\rho_0=2\times10^{-4}$,
    $\alpha=0.1$).  Line identity: black dashed $=$ Kerr NTP (coincides
    with blue dotted $=$ viscous Kerr); red dash-dotted $=$
    NTP Hayward+DE (coincides with green solid $=$ viscous Hayward+DE)
    in the NTP panel --- both overlaps are exact (see text).
    \textbf{(a)}~Full disk; \textbf{(b)}~inner disk zoomed to
    $r/M_{\BH}\leq15$, where the geometric and viscous modifications
    are most pronounced.}
  \label{fig:flux}
\end{figure}

Figure~\ref{fig:flux} shows the dimensionless flux $\Fcal^*\equiv
M_{\BH}^2\Fcal$ for the four cases we have been discussing.  Panel~(a)
covers the whole disk out to $r/M_{\BH}=35$; panel~(b) zooms into
the inner region $r/M_{\BH}\leq15$, where the Hayward core and the DE
field actually leave their mark.  Line styles are: Kerr NTP (black
dashed), viscous Kerr (blue dotted), NTP Hayward+DE (red
dash-dotted), viscous Hayward+DE (green solid).

The Hayward core on its own pulls $\rISCO$ in a little, from
$4.614\,M_{\BH}$ to $4.543\,M_{\BH}$, which narrows the flux peak
and lifts it by about $4\,\%$.  Adding the DE field pushes the
enclosed mass above $M_{\BH}$ at every radius, and the net effect is
a slightly outward ISCO, at $4.683\,M_{\BH}$ -- but the binding
energy there is considerably deeper, which is what drives
$\eta$ up to $8.71\,\%$.  The viscous correction (green solid) then
adds a further boost to the inner-disk flux through the non-zero
torque $\Tcal_{\rm in}$.

Only two curves are really visible in panel~(a), because
$\Fcal_{\rm NTP}$ does not care about $\alpha$ at all -- it only
depends on the geodesics -- so the Kerr NTP and viscous Kerr curves
sit exactly on top of each other, and likewise for the two
Hayward+DE curves.  Zooming in, as panel~(b) does, is what actually
lets the viscosity enhancement show up, in the region
$r/M_{\BH}\lesssim7$.

\begin{figure}[ht]
  \centering
  \includegraphics[width=\columnwidth]{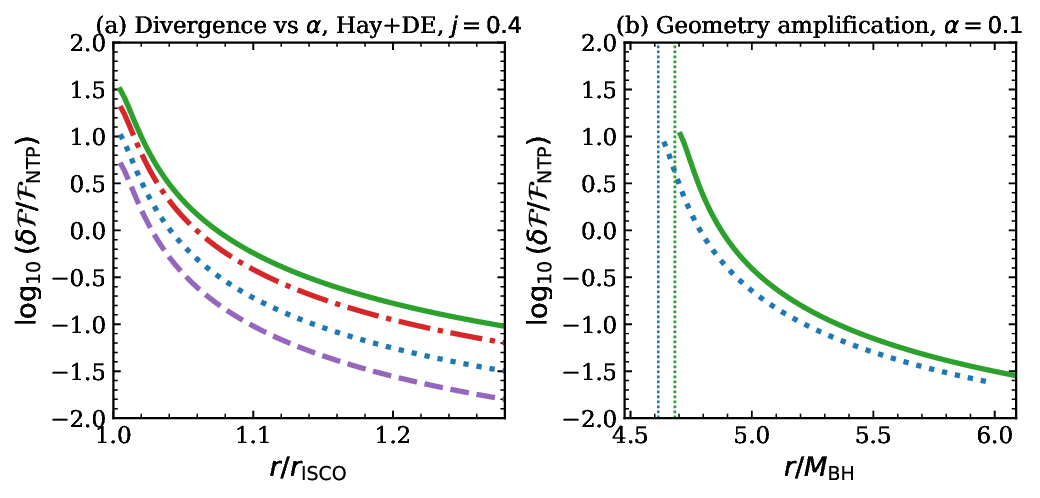}
    \caption{Inner-torque divergence shown on logarithmic scale
    ($y$-axis: $\log_{10}(\delta\mathcal{F}/\mathcal{F}_{\rm NTP})$),
    directly visualising the analytic result $\delta\mathcal{F}/
    \mathcal{F}_{\rm NTP}=\mathcal{T}_{\rm in}/I(r)\to\infty$ as
    $r\to\rISCO^+$ [Eq.~(\ref{eq:dF})].
    \textbf{(a)}~$\log_{10}(\delta\mathcal{F}/\mathcal{F}_{\rm NTP})$
    vs.\ $r/\rISCO$ for $\alpha=0.05$ (purple dashed), $0.1$ (blue
    dotted), $0.2$ (red dash-dotted), $0.3$ (green solid), Hayward+DE
    background, $j=0.4$: all curves diverge at the left boundary
    ($r\to\rISCO^+$) with amplitude scaling monotonically with
    $\alpha$, from $\alpha=0.05$ (bottom) to $\alpha=0.3$ (top).
    \textbf{(b)}~Same quantity at fixed $\alpha=0.1$: viscous Kerr
    (blue dotted, $\rISCO/M_{\BH}=4.614$, vertical dotted line)
    vs.\ viscous Hayward+DE (green solid, $\rISCO/M_{\BH}=4.683$,
    left vertical line).  The geometric modification raises
    $\mathcal{T}_{\rm in}$ at $\rISCO$, amplifying the divergence
    amplitude relative to vacuum Kerr.}
  \label{fig:divergence}
\end{figure}

Figure~\ref{fig:divergence} gives a more direct look at the
divergence itself -- the analytic result
$\delta\Fcal/\Fcal_{\rm NTP}=\Tcal_{\rm in}/I(r)\to\infty$ as
$r\to\rISCO^+$ [Eq.~(\ref{eq:dF})] -- and at how much the modified
geometry amplifies it.

Panel~(a) plots $\log_{10}(\delta\Fcal/\Fcal_{\rm NTP})$
against $r/\rISCO$ for $\alpha=0.05$ (purple dashed), $0.1$ (blue
dotted), $0.2$ (red dash-dotted), $0.3$ (green solid), all with the
Hayward+DE background: every curve climbs as $r\to\rISCO^+$, with a
size set by $\alpha$.  Panel~(b) puts viscous Kerr (blue dotted,
$\rISCO=4.614\,M_{\BH}$) side by side with viscous Hayward+DE (green
solid, $\rISCO=4.683\,M_{\BH}$) at a fixed $\alpha=0.1$.  The two
curves separate because the modified geometry changes the boundary
pressure $P_{\rm tot}(\rISCO)$ and scale height $H(\rISCO)$ that feed
into $\Tcal_{\rm in}$, as discussed in Sect.~\ref{sec:disk}.

\subsection{Radiative efficiency and how it decouples from viscosity}
\label{sec:res_eta}

\begin{figure}[ht]
  \centering
  \includegraphics[width=\columnwidth]{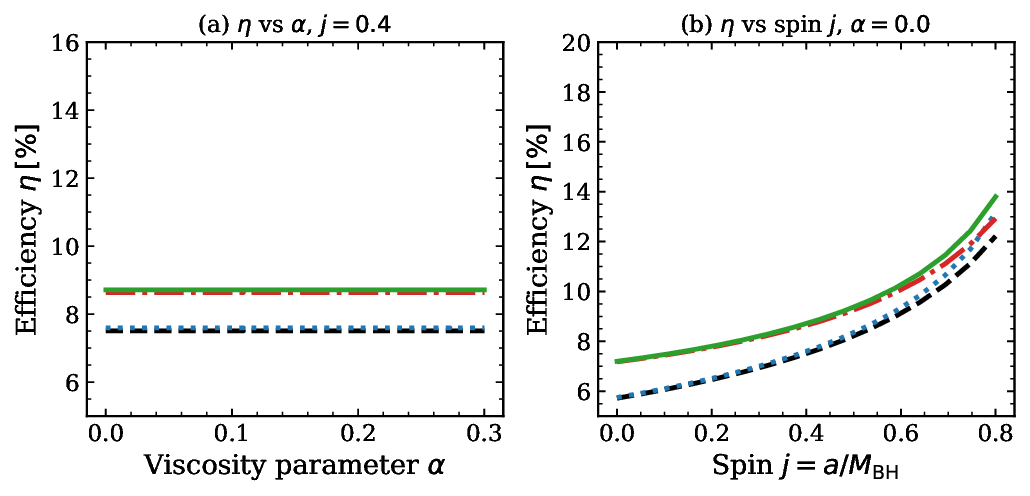}
  \caption{Bolometric efficiency $\eta\,[\%]$ for the four
    configurations: vacuum Kerr (black dashed), Hayward only (blue
    dotted), DE only (red dash-dotted), Hayward+DE (green solid).
    \textbf{(a)}~$\eta$ vs.\ viscosity parameter $\alpha$ at $j=0.4$:
    all curves are perfectly flat, confirming $\eta$ is strictly
    independent of $\alpha$.
    \textbf{(b)}~$\eta$ vs.\ spin $j$ at $\alpha=0$: the Hayward+DE
    model (green) systematically exceeds vacuum Kerr (black) at all
    spin values, reaching $\eta\approx12\,\%$ at $j\approx0.7$.}
  \label{fig:efficiency}
\end{figure}

Figure~\ref{fig:efficiency} shows how $\eta$ behaves for the four
configurations introduced above.  Plotted against $\alpha$, all four
curves are perfectly flat, which is just the analytic result of
Sect.~\ref{sec:observables} confirmed numerically to machine
precision.  Plotted against spin instead, at $\alpha=0$, the
Hayward+DE curve (green solid) sits above vacuum Kerr (black dashed)
at every spin, and reaches $\eta\approx12\,\%$ already at
$j\approx0.7$ -- a value that vacuum Kerr alone only reaches once
$j>0.9$~\cite{DavisLaor2011}.

This is what makes a two-step measurement strategy possible.
Because $\eta$ knows nothing about $\alpha$, a bolometric measurement
of $\eta$ constrains the geometry -- $j$, $l$, and $\rho_0$ -- on its
own.  Once those are fixed, the high-frequency spectral ratio in
Fig.~\ref{fig:spectra}(b) can then be used to pin down $\alpha$
separately, without the two ever getting tangled up with each other.

\subsection{Spectral signatures}
\label{sec:res_spec}

\begin{figure}[ht]
  \centering
  \includegraphics[width=\columnwidth]{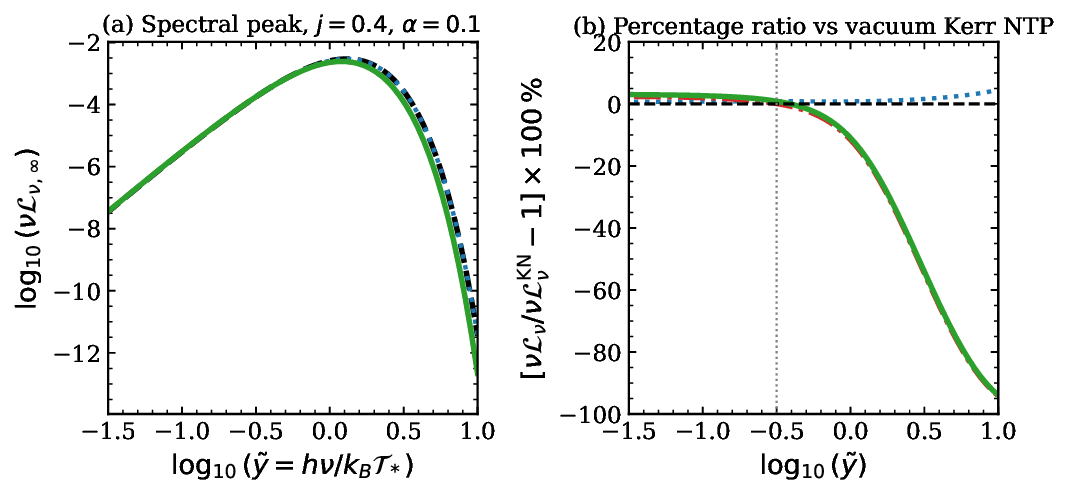}
  \caption{Spectral observables at $j=0.4$, $f_{\rm col}=1.7$,
    $\alpha=0.1$.  Line identity: black dashed $=$ Kerr NTP, blue
    dotted $=$ viscous Kerr, red dash-dotted $=$ NTP Hayward+DE,
    green solid $=$ viscous Hayward+DE.
    \textbf{(a)}~Absolute spectral luminosity $\log_{10}(\nu\Lcal_\nu)$
    near the spectral peak: the Hayward core raises the peak slightly;
    the DE contribution broadens and cools the disk, lowering the
    bulk luminosity of the combined geometry despite its higher
    $\eta$ (Table~\ref{tab:isco}).
    \textbf{(b)}~Percentage deviation
    $[\nu\Lcal_\nu/\nu\Lcal_\nu^{\rm KN}-1]\times100\,\%$ relative to
    vacuum Kerr NTP (black dashed at 0), shown honestly over the full
    range: the Hayward+DE geometry is systematically fainter in bulk
    (down to $-90\,\%$), with the viscosity correction (green solid)
    providing partial compensation only at high frequency
    ($\log_{10}\tilde{y}\gtrsim0.5$).}
  \label{fig:spectra}
\end{figure}

Figure~\ref{fig:spectra} turns to the spectrum itself.  Panel~(a)
shows the absolute spectral luminosity near the peak,
$\log_{10}(\nu\Lcal_{\nu,\infty})$, for the same four cases.  The
Hayward core alone lifts the peak a little, since it pulls the ISCO
in; the DE field does the opposite to the bulk spectrum -- it
broadens the disk and cools it down, so the overall luminosity drops
even though $\eta$ has gone up.  There is no contradiction here:
$\eta$ only measures what fraction of the rest-mass energy comes out
as radiation at the ISCO, while the bulk luminosity is an integral
over the whole disk, which in the combined geometry happens to be
wider and cooler.

Panel~(b) makes this quantitative, plotting the percentage deviation
$[\nu\Lcal_\nu/\nu\Lcal_\nu^{\rm KN}-1]\times100\,\%$ relative to
vacuum Kerr NTP over the full frequency range, without hiding the
parts that look unflattering.  The Hayward+DE curves sit below vacuum
Kerr for most of the spectrum, by as much as $90\,\%$ at low
$\tilde{y}$, which is just the broader and cooler disk again.  Only
the viscous correction (green solid, $\alpha=0.1$) claws some of this
back, and only near the high-frequency tail
($\log_{10}\tilde{y}\gtrsim0.5$), where the divergence right at the
ISCO (Fig.~\ref{fig:divergence}) starts to matter.  We think it is
worth being upfront that this bulk suppression is real and not an
artefact -- it is a separate effect from the local inner-torque
divergence discussed above, which remains a clean signature near
$\rISCO$ regardless of what the bulk spectrum is doing.

\begin{figure}[ht]
  \centering
  \includegraphics[width=\columnwidth]{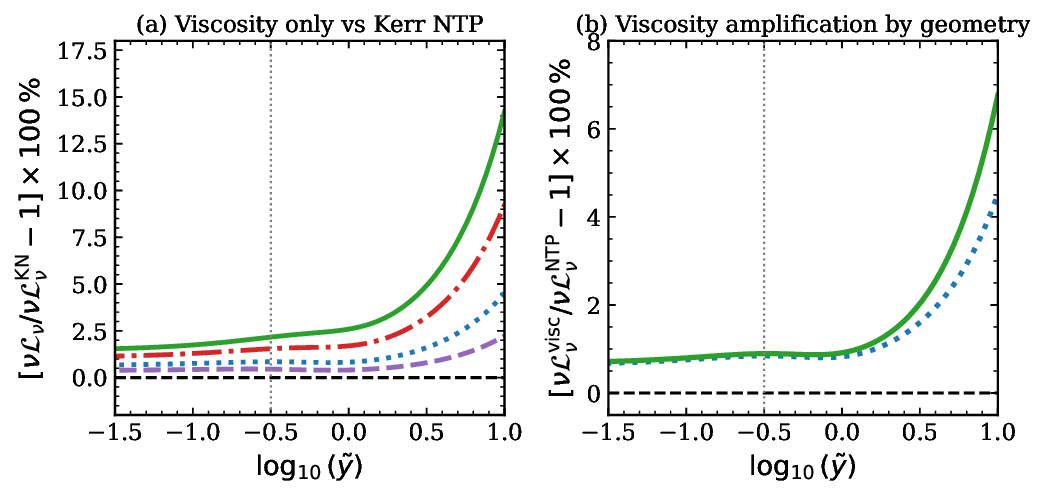}
  \caption{Percentage viscosity amplification, isolating the
    non-zero inner torque from the bulk spectral normalisation.
    \textbf{(a)}~Deviation from Kerr NTP for the viscosity effect
    alone ($l=0$, $\rho_0=0$) at $\alpha=0.05$ (purple dashed),
    $0.1$ (blue dotted), $0.2$ (red dash-dotted), $0.3$ (green
    solid): the excess grows monotonically with $\alpha$, reaching
    $\approx14\,\%$ at $\alpha=0.3$.
    \textbf{(b)}~Viscosity amplification ratio
    $[\nu\Lcal_\nu^{\rm visc}/\nu\Lcal_\nu^{\rm NTP}-1]\times100\,\%$
    computed on the \emph{same} background, comparing vacuum Kerr
    (blue dotted, $0.80$--$4.52\,\%$) and Hayward+DE (green solid,
    $0.87$--$6.76\,\%$) at $\alpha=0.1$: the modified geometry
    systematically amplifies the viscosity correction, providing a
    clean, monotonic, and unambiguously positive observational
    signature.}
  \label{fig:amplification}
\end{figure}

To separate the viscosity effect cleanly from the bulk spectral
changes just described, Fig.~\ref{fig:amplification} looks at ratios
taken on a common background rather than against vacuum Kerr NTP,
and this is where we think the strongest observational signature of
the whole model sits.  Panel~(a) shows the percentage excess from
viscosity alone, in vacuum Kerr ($l=0$, $\rho_0=0$), for
$\alpha=0.05$--$0.3$: it grows steadily with $\alpha$, from around
$2\,\%$ up to about $14\,\%$ at $\log_{10}\tilde{y}=1$.

Panel~(b) is the key comparison: the viscosity amplification ratio
$[\nu\Lcal_\nu^{\rm visc}/\nu\Lcal_\nu^{\rm NTP}-1]\times100\,\%$,
computed separately for vacuum Kerr (blue dotted) and for Hayward+DE
(green solid), each against its own NTP baseline, at a fixed
$\alpha=0.1$.  Because both curves use their own background as the
reference point, whatever is different between them can only come
from the inner torque itself, not from the overall brightness or
shape of the disk.  Over $\log_{10}\tilde{y}\in[-0.5,+1]$ the
Hayward+DE curve runs consistently above the Kerr one -- $0.87$ to
$6.76\,\%$ against $0.80$ to $4.52\,\%$ -- which is exactly the
amplification of $\Tcal_{\rm in}$ we argued for in
Sect.~\ref{sec:disk}.  Unlike the bulk spectral ratio in
Fig.~\ref{fig:spectra}(b), this one stays positive and grows smoothly
throughout, and we regard it as the most dependable observational
handle this model offers.

\subsection{How this compares with what is actually observed}
\label{sec:obs}

At $j=0.4$ the model gives $\eta=8.71\,\%$ against $7.51\,\%$ for
vacuum Kerr at the same spin -- a $16\,\%$ increase without needing
any extra rotation.  Push the spin to $j=0.7$ and $\eta$ reaches
about $12\,\%$, which lands squarely in the range Davis \&
Laor~\cite{DavisLaor2011} infer for luminous quasars.  A good part of
their sample needs $\eta>7.5\,\%$, something vacuum Kerr cannot
supply below $j\approx0.4$ but that our model reaches comfortably at
moderate spin.  We would treat these as order-of-magnitude
comparisons rather than a fit to data -- doing this properly would
mean folding in radiative transfer, Comptonization, and inclination,
none of which we have attempted here.

The enhanced dissipation near the inner edge also offers a plausible
route to the soft X-ray excess seen in around half of Seyfert~1
galaxies~\cite{Done2012}: the amplified torque raises the local
effective temperature ($T_{\rm eff}\propto\Fcal^{1/4}$) and hardens
the continuum below about $2\,{\rm keV}$, without having to invoke a
separate warm corona.  Testing this properly against real Seyfert
spectra, with something like an \textsc{xspec} \texttt{diskbb} model,
is left for later work.

Finally, the same divergence that drives the flux up near $\rISCO$
(Fig.~\ref{fig:divergence}) also pushes up the local radiation
pressure, and disks dominated by radiation pressure are known to be
prone to the Lightman--Eardley secular
instability~\cite{LightmanEardley1974}.  If that instability is
triggered here, one might expect quasi-periodic oscillations with a
frequency shifted by something like
$(3/2)|\delta\rISCO|/\rISCO\sim1\,\%$ relative to vacuum Kerr.  We
want to be clear that this is speculation rather than a result: a
proper answer would need a time-dependent stability analysis, which
we have not carried out.

\section{Conclusions}
\label{sec:conclusions}

In this paper we have worked out what happens to a viscous accretion
disk once the stress-free inner boundary of the Novikov--Thorne--Page
model is dropped, for a rotating Hayward black hole sitting in a
Kiselev quintessence field.  We have tried to be honest about the
spacetime itself: the static Hayward core and the static quintessence
field are each exact solutions of Einstein's equations on their own,
with well-defined stress-energy tensors, but rotating and combining
them the way we have -- inserting a running mass into the Kerr form
-- is standard practice in this literature rather than a rigorous
derivation, and following Kamenshchik \& Petriakova~\cite{Kamenshchik2023}
we expect small violations of the field equations close to the core
even though the metric returns to exact Kerr far away.  A genuine
first-principles construction, whether through the modified
Newman--Janis procedure of Azreg-A\"inou or by solving the coupled
field equations directly, would close this gap and seems worth doing
in future work. Within that spacetime, though, the disk physics
follows through cleanly: the surface density in
Eq.~(\ref{eq:Sigma}) stays positive everywhere because the sign of
the boundary torque is fixed by causality, not by hand, and the
resulting flux correction $\delta\Fcal/\Fcal_{\rm NTP}$ genuinely
diverges as $r\to\rISCO^+$, with the modified geometry making that
divergence larger rather than smaller.

The result we find most useful is that the radiative efficiency
$\eta$ turns out to depend only on the geodesic structure of the
spacetime and not at all on the viscosity parameter $\alpha$ -- a
statement we can prove analytically and which holds numerically to
better than $10^{-8}\,\%$.  At $j=0.4$ this alone lifts $\eta$ from
$7.51\,\%$ in vacuum Kerr to $8.71\,\%$ once the Hayward core and
the DE field are included, and because $\alpha$ never enters, a
bolometric measurement of $\eta$ can constrain the geometry on its
own, leaving the high-frequency spectral ratio free to constrain
$\alpha$ separately.  That ratio is not the cleanest quantity if
compared directly against vacuum Kerr, since the bulk spectrum of
the Hayward+DE disk is genuinely fainter there; but when the
viscosity correction is compared against its own geometry's NTP
baseline instead, it comes out larger for Hayward+DE than for vacuum
Kerr across the whole observable band -- $0.87$ to $6.76\,\%$
against $0.80$ to $4.52\,\%$ at $\alpha=0.1$ -- which we think is
the most dependable signature this model has to offer, and one that
instruments like \textit{NICER} and \textit{NuSTAR} should be able to
probe.  Looking ahead, replacing the $\alpha$ prescription with full
GRMHD simulations, feeding Eq.~(\ref{eq:flux}) into \textsc{xspec}
for actual spectral fitting, and working out the vertical epicyclic
frequency exactly rather than approximately would all be natural next
steps.

\section*{Declaration of Generative AI and AI-assisted technologies in the writing process}
During the preparation of this work, the author did not use any generative AI or AI-assisted technologies in the research and concept. The author take full responsibility for the content of the publication. The author use AI assistant for structuring the manuscript only in latex.
\section*{Declaration of competing interest}
The author declares that they have no known competing financial interests or personal relationships that could have appeared to influence the work reported in this paper.
\section*{Data Availability Statement}
This manuscript has no associated data or the data will not be deposited. [Authors' comment: This is a purely theoretical and mathematical physics study. All numerical results, curves, and physical quantities presented in the figures were generated directly using the analytical equations, initial conditions, and numerical integration methodologies explicitly detailed in the text. The corresponding Python scripts used to compile the data and generate the plots are available from the single corresponding author upon reasonable request.]

\section*{Acknowledgement}
	The author, Sandip Dutta, wishes to express sincere gratitude to the Department of Applied Mathematics at the Dinabandhu Andrews Institute of Technology and Management (DAITM) for providing the academic environment and computational facilities necessary to conduct this theoretical research. The author also thanks Dr. Ritabrata Biswas for his guidance and inspiration to the work.

\appendix
\section{Discriminant Condition and ISCO Validity}
\label{app:disc}

The discriminant $D$ in Eq.~(\ref{eq:Omega}) dictates whether real
circular orbits exist.  For the running-mass metric Eq.~(\ref{eq:Mtot}), Eq.~(\ref{eq:Dform}) gives
\begin{equation}
\label{eq:Dform}
  D = \frac{4[M(r)-rM'(r)]}{r}, \quad
  M'(r) = \frac{dM_H}{dr} + \frac{dM_{\DE}}{dr},
\end{equation}
where Eq.~(\ref{eq:Mprime}) gives
\begin{equation}
\label{eq:Mprime}
  \frac{dM_H}{dr} = \frac{3M_{\BH}l^3 r^2}{(r^3+l^3)^2},\quad
  \frac{dM_{\DE}}{dr} = 4\pi r^2\rho_{\DE}(r).
\end{equation}
Circular orbits are valid ($D>0$) when $M(r)>rM'(r)$, i.e., the
enclosed mass exceeds the local mass gradient.  For our benchmark
parameters ($l=0.5M_{\BH}$, $\rho_0=2\times10^{-4}$), we have
verified numerically that $D>0$ throughout
$r\in[r_g,\,40M_{\BH}]$, confirming orbital stability.  The ISCO
condition $dL/dr=0$ is equivalent to the marginal stability of the
radial effective potential and to $\kappa^2=0$ (vanishing radial
epicyclic frequency) for any stationary axisymmetric
spacetime~\cite{Bardeen1972}, provided $D>0$.


\end{document}